\documentstyle[12pt,epsf]{article}

\newcommand{\noi}{\noindent}
\newcommand{\eq}{\begin{equation}}
\newcommand{\en}{\end{equation}}
\newcommand{\eqa}{\begin{eqnarray}}
\newcommand{\ena}{\end{eqnarray}}

\newcommand{\vp}{{\vec p}}

\newcommand{\vR}{{\vec R}}
\newcommand{\vzero}{{\vec 0}}

\newcommand{\qr}{q^{\prime}}

\newcommand{\cak}{ {\cal K} }

\newcommand{\inta}{\int_{A}~}

\newcommand{\aleq}{\mbox{}_{\textstyle \sim}^{\textstyle < }}

\newcommand{\ra}{\rightarrow}

\newcommand{\lra}{\longrightarrow}

\begin{document}

\hbox{}
\noindent {21 April 1997} \hfill JINR E2-97-142 \\
\mbox{} \hfill FSU-SCRI-97-37   \\
\mbox{} \hfill IFUP--TH 14/97   \\

\vspace*{1.0cm}

\begin{center}
\renewcommand{\thefootnote}{\fnsymbol{footnote}}
\setcounter{footnote}{0}
{\LARGE  The Coulomb law in the pure gauge $U(1)$
theory on a lattice}\footnote{
Work supported by the Deutsche Forschungsgemeinschaft under research 
grant Mu 932/1-4 and EEC--contract CHRX-CT92-0051
and by the US DOE under grants
\#~DE-FG05-85ER250000 and \#~DE-FG05-96ER40979.
} \\

\vspace*{1.0cm}
{\large
G.~Cella$\mbox{}^1$,
U.M.~Heller$\mbox{}^2$,
V.K.~Mitrjushkin$\mbox{}^{3}$\footnote{present address: BLTPh,
                                       JINR, Dubna, Russia}
and
A.~Vicer\'e$\mbox{}^1$
}\\
\vspace*{0.7cm}
{\normalsize
$\mbox{}^1$ {\em INFN in Pisa and Dipartimento di Fisica dell'Universit\'a
di Pisa,  Italy }\\
$\mbox{}^2$ {\em SCRI, Florida State University, Tallahassee,
FL 32306-4052, USA} \\
$\mbox{}^3$ {\em Humboldt-Universit\"{a}t zu Berlin, Institut f\"{u}r Physik,
10115 Berlin, Germany}
}\\
\vspace*{1cm}

{\bf Abstract}
\end{center}

We study the heavy charge potential in the  Coulomb phase of pure gauge
compact $U(1)$ theory on the lattice. We calculate the static potential
$V_W(T,{\vec R})$ from Wilson loops on a $16^3 \times 32$ lattice and
compare with the predictions of lattice perturbation theory. We investigate
finite size effects and, in particular, the importance of non--Coulomb
contributions to the potential. We also comment on the existence of a
maximal coupling in the Coulomb phase of pure gauge $U(1)$ theory.

\section{Introduction}

In 1974 Wilson proposed the compact lattice formulation of pure gauge 
$U(1)$ theory \cite{wil}.  The action of this model is 

\eq
S_W(U ) = \beta \sum_x \sum_{\mu > \nu }
         \Bigl( 1 - \cos \theta_{x,\mu\nu} \Bigr) ~,
\qquad \beta =\frac{1}{g^2}~,
                                              \label{wil_action}
\en

\noi where $~g^2 ~$ is the bare coupling constant, 
and the link variables are $U_{x\mu} =\exp (i\theta_{x\mu})$,
$\theta_{x \mu} \in (-\pi, \pi]$.  The plaquette angles are given by
$\theta_{x,\, \mu \nu} = \theta_{x,\, \mu} + \theta_{x + \hat{\mu},\, \nu}
- \theta_{x + \hat{\nu},\, \mu} - \theta_{x,\, \nu}$.
This action makes up the pure gauge part of the full QED action $S_{QED}$,
which is supposed to be compact if we consider QED as arising from a
subgroup of a non--abelian (e.g., grand unified) gauge theory \cite{pol}.

\noi The average of any gauge--invariant functional ${\cal O}(U)$
is

\eq
\langle {\cal O}(U) \rangle = Z^{-1} \int \! \prod_{x\mu} \! dU_{x \mu} \,
{\cal O}(U) \cdot e^{- S_W(U)}~,
\qquad dU_{x\mu} =\frac{d\theta_{x\mu}}{2\pi}~,
                                          \label{func}
\en

\noi where $Z$ is the partition function, 
defined formally by $\langle 1 \rangle = 1$.
At small enough $~\beta$'s the strong coupling expansion shows an
area--law behaviour of the Wilson loops, while at large $~\beta$'s a
deconfined phase exists. The two phases are separated by a phase
transition at some `critical' value $\beta^\ast$ whose existence was
assumed in \cite{wil}.  The weak coupling -- deconfined -- phase is
expected to be a Coulomb phase, {\it i.e.,} a phase with massless
noninteracting vector bosons (photons) and Coulomb--like interactions
between static charges.

There are scarcely any doubts about the existence of the Coulomb phase in the
weak coupling region for Wilson's QED, though -- to our knowledge --
there is no rigorous proof.
It is worthwhile to mention that for the Villain approximation \cite{vil}
such a proof is available \cite{guth,sei}.
However, a detailed study has shown that the Villain action
is quantitatively a rather bad approximation to the Wilson action in
the weak coupling region \cite{jk}.

A perturbative expansion suggests for the lattice potential 
$V^{latt} (\vR )$ the expression 

\eq
V^{latt} (\vR ) = \sum_{n=1}^{\infty} g^{2n} \cdot V_n(\vR )~,
\en

\noi where up to an additive constant $~V_1~$ is the 
lattice analog of the 

continuum Coulomb potential $~\sim 1/4\pi R~$ 
(for an explicit expression see (\ref{v_coul_latt_fin}) below).
One--loop corrections  ($\sim g^4$) do not change the
functional dependence of the potential but at the two--loop level ($\sim
g^6$) non--Coulomb--like contributions appear (see below). The analytical
and numerical study of these contributions, as well as of the finite volume
behavior of the potential, constitutes the aim of the present work.

The behavior of the heavy charge potential in the weak coupling phase in
pure $U(1)$ gauge theory was the subject of several numerical studies (see,
e.g.  \cite{bhan,dgt1,jnz,sw1}).  In all cases consistency with a
Coulomb--like behavior was reported. However, we feel that a more elaborate
and systematic study is necessary. In particular, Monte Carlo simulations
give precise enough measurements of the potential, that finite size effects
and finite lattice spacing effects, {\it i.e.,} deviations of the lattice
Coulomb potential from the continuum $1/R$ behavior
should not be neglected. In addition, finite $T$ effects in the extraction
of the potential from Wilson loops have to be taken into account. We shall
find, on the other hand, that the non-Coulomb--like contributions to the
potential are negligible.

We consider this systematic study of the Coulomb phase in pure gauge
$U(1)$ theory a necessary step before attempting an investigation of
lattice QED with fermions with the aim of obtaining a continuum limit
that reproduces weak coupling continuum QED.

The second section is devoted to the two-loop perturbative study of the
potential, $V_P(\vR)$, as extracted from Polyakov loop correlations.
Special emphasis has been put on the study of the {\it non}--Coulomb--like
contributions that appear at two--loop order, {\it i.e.,} at order
$O(g^6)$. In the third section we discuss the potential, $V_W(T,\vR)$, as
extracted from Wilson loops. We present the results of Monte Carlo
simulations and their comparison with perturbation theory. Indeed,
perturbation theory will play a major role in our extraction of a
renormalized coupling, $g_R^2$, from the numerical simulations.
In section \ref{max_coup} we address the question of the existence of a
maximal coupling in pure gauge U(1).
The last section is reserved for conclusions and discussions.

\section{The potential from Polyakov loop correlations at $O(g^6)$}

One way to define the heavy charge potential
$V =V_P(\vR)$ is

\eq
V_{P}(\vR) = -\frac{1}{N_4} \cdot \ln \Gamma_{P}(\vR)~,
                                      \label{v_p1}
\en

\noi where we consider a finite lattice of size $N_s^3 \times N_4$ and
$\Gamma_{P}(\vR)$ is the Polyakov loop correlator 

\eq
\Gamma_P({\vec R}) = \frac{1}{Z} \inta 
e^{ig\sum_{x\mu}J^P_{x\mu}A_{x\mu}} \cdot e^{-S_W(A)}~,
\en

\noi if the currents $J^P_{x\mu}$ correspond to a static heavy 
charge--anticharge pair   
\eqa
J^P_4(x) &=& \left\{ \begin{array}{ll}
 \delta_{{\vec x},{\vec 0}}-\delta_{{\vec x},{\vec R}} 
& \quad \mbox{at} \quad 0 \leq x_4 \leq N_4 -1 \\
0 & \quad \mbox{otherwise}
\end{array}
\right. ~,
\ena

\noi and $J^P_i(x)=0$, $i=1,2,3$.

\begin{figure}
\begin{center}
\vspace{-1.0cm}
\setlength{\unitlength}{0.75cm}
\begin{picture}(17,3.5)
\thicklines
%
\put(1,1){\circle{0.2}}
\put(4,1){\circle{0.2}}
\put(1.1,1){\line(1,0){2.8}}
\put(0.5,0.9){$g$}
\put(4.3,0.9){$g$}
\put(1.0,1.5){a)}
%
\put(7,1){\circle{0.2}}
\put(10,1){\circle{0.2}}
\put(8.5,1){\circle*{0.2}}
\put(7.1,1){\line(1,0){2.8}}
\put(8.5,1.5){\oval(0.6,0.6)[t]}
\put(8.5,1){\line(3,5){0.3}}
\put(8.2,1.5){\line(3,-5){0.3}}
\put(6.5,0.9){$g$}
\put(10.3,0.9){$g$}
\put(8.5,0.5){$g^2$}
\put(7.0,1.5){b)}
%
\put(13,1){\circle{0.2}}
\put(16,1){\circle{0.2}}
\put(14.0,1){\circle*{0.2}}
\put(15.0,1){\circle*{0.2}}
\put(13.1,1){\line(1,0){2.8}}
\put(14.0,1.5){\oval(0.6,0.6)[t]}
\put(14.0,1){\line(3,5){0.3}}
\put(13.7,1.5){\line(3,-5){0.3}}
\put(15.0,1.5){\oval(0.6,0.6)[t]}
\put(15.0,1){\line(3,5){0.3}}
\put(14.7,1.5){\line(3,-5){0.3}}
\put(12.5,0.9){$g$}
\put(16.3,0.9){$g$}
\put(14.0,0.5){$g^2$}
\put(15.0,0.5){$g^2$}
\put(13.0,1.5){c)}
\end{picture}

%
\unitlength 0.75cm
\begin{picture}(17,3.5)
\thicklines
%
\put(1,1){\circle{0.2}}
\put(4,1){\circle{0.2}}
\put(2.5,1){\circle*{0.2}}
\put(1.1,1){\line(1,0){2.8}}
\put(2.5,1.5){\oval(0.6,0.6)[t]}
\put(2.5,1){\line(3,5){0.3}}
\put(2.2,1.5){\line(3,-5){0.3}}
\put(2.5,1.8){\circle*{0.2}}
\put(2.5,2.3){\oval(0.6,0.6)[t]}
\put(2.5,1.8){\line(3,5){0.3}}
\put(2.2,2.3){\line(3,-5){0.3}}
\put(0.5,0.9){$g$}
\put(4.3,0.9){$g$}
\put(2.5,0.5){$g^2$}
\put(3.0,1.5){$g^2$}
\put(1.0,1.5){d)}

\put(7,1){\circle{0.2}}
\put(10,1){\circle{0.2}}
\put(8.5,1){\circle*{0.2}}
\put(7.1,1){\line(1,0){2.8}}
\put(8.5,1.5){\oval(0.6,0.6)[t]}
\put(8.5,1){\line(3,5){0.3}}
\put(8.2,1.5){\line(3,-5){0.3}}
\put(8.5,0.5){\oval(0.6,0.6)[b]}
\put(8.2,0.5){\line(3,5){0.3}}
\put(8.5,1.0){\line(3,-5){0.3}}
\put(6.5,0.9){$g$}
\put(10.3,0.9){$g$}
\put(9.0,0.5){$g^4$}
\put(7.0,1.5){e)}
%
\put(13,1.5){\circle{0.2}}
\put(13,0.5){\circle{0.2}}
\put(16,1.5){\circle{0.2}}
\put(16,0.5){\circle{0.2}}
\put(14.5,1){\circle*{0.2}}
\put(13.0949,1.4684){\line(3,-1){2.8}} 
\put(13.0949,0.5316){\line(3,1){2.8}}
\put(12.5,1.45){$g$}
\put(12.5,0.45){$g$}
\put(16.3,1.45){$g$}
\put(16.3,0.45){$g$}
\put(14.5,0.4){$g^2$}
\put(14.5,1.5){f)}
\end{picture}

%
\unitlength 0.75cm
\begin{picture}(17,3.5)
\thicklines
\put(1,1){\circle{0.2}}
\put(4,1){\circle{0.2}}
\put(2.0,1){\circle*{0.2}}
\put(3.0,1){\circle*{0.2}}
\put(1.1,1){\line(1,0){2.8}}
\put(2.5,1.0){\oval(1.0,1.3)[t]}
\put(2.5,1.0){\oval(1.0,1.3)[b]}
\put(0.5,0.9){$g$}
\put(4.3,0.9){$g$}
\put(1.5,0.4){$g^2$}
\put(3.1,0.4){$g^2$}
\put(1.0,1.5){g)}
\end{picture}
\end{center}
\caption{Different diagrams contributing to the potential $V_P(\vR )$.}
\label{fig:diagrams}
\end{figure}
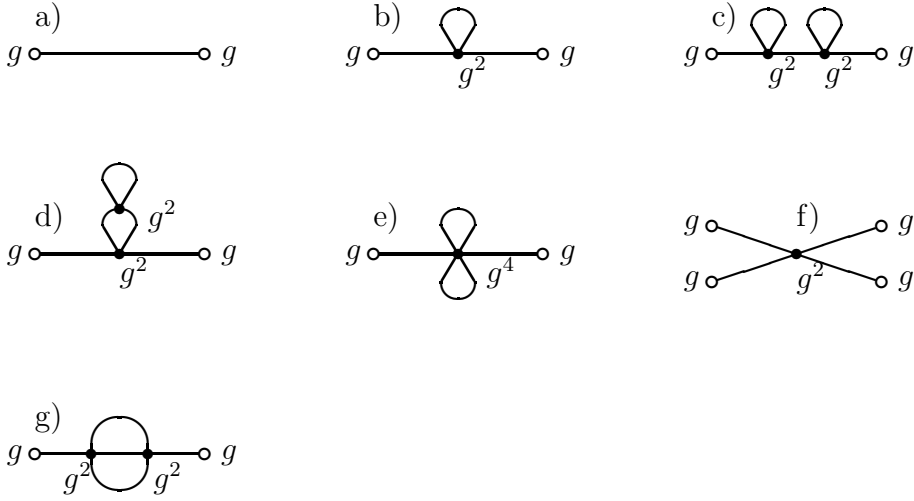

We calculated the potential $V_P(\vR)$ perturbatively up to order $O(g^6)$.
Graphically the different contributions are shown in Figure
\ref{fig:diagrams}.

%

In the tree approximation (Figure \ref{fig:diagrams}a) one obtains
the lattice Coulomb potential

\eq
g^2 V_{Coul}^{(N_s)}(\vR ) = \frac{g^2}{N_s^3} \sum_{\vp \ne 0} 
\frac{1-\cos \vp\vR} {{\vec \cak}^{\, 2}  }~,
                \label{v_coul_latt_fin}
\en

\noi where $\cak_{i} =2\sin \frac{p_{i}}{2}$.
In the infinite volume limit, $~N_s \to \infty~$, this potential becomes

\eq
g^2 V_{Coul}^{(\infty)}(\vR) = \frac{g^2}{(2\pi )^3}
\int_{-\pi}^{\pi} \! d{\vec p} ~ \frac{1 - \cos \vp\vR }
{{\vec \cak}^{\, 2}  }~,
\quad N_s \lra \infty~,
                \label{v_coul_latt_infin}
\en

\noi The continuum analog of this expression is 

\eq
g^2 V_{cont}({\vec R}) = -\frac{g^2}{4\pi R} + \mbox{Const} ~.
                \label{v_coul_cont}
\en

\noi The lattice Coulomb potential defined in 
eq.~(\ref{v_coul_latt_fin}) becomes close to the continuum 
expression in eq.~(\ref{v_coul_cont}) when 
$1 \ll R \ll \frac{1}{2}N_s~$
for sufficiently large lattice size $N_s$.
For lattice sizes that are
typically used in numerical simulations (i.e. $N_s \sim 16 \div 32$)
the lattice Coulomb potential in eq.~(\ref{v_coul_latt_fin})
differs considerably from the continuum expression in
eq.~(\ref{v_coul_cont}).

The $O(g^4)$ contribution (Figure \ref{fig:diagrams}b) as
well as $O(g^6)$ contributions shown in 
Figures \ref{fig:diagrams}c to \ref{fig:diagrams}e
result in a renormalization of the coupling without
changing the form of the potential.
However, two diagrams at order $\sim O(g^6)$
contain non--Coulomb--like contributions.
One of them is the four--prong--spider graph 
(Figure \ref{fig:diagrams}f) which gives the contribution
to the potential

\eqa
V^{(f)}(\vR) &=& - \frac{2g^6}{3N_s^9} \sum_{\{ \vp^{(i)} \} } 
\delta_{\vp^{(1)} + \ldots + \vp^{(4)},0} 
\prod_{i=1}^{4} \sin \frac{ \vp^{(i)} \vR }{2} 
\cdot  \sum_{j=1}^3 \prod_{i=1}^{4}
  \frac{\cak_{j}(p^{(i)}) }{{\vec \cak}^2(p^{(i)}) } ~.
\label{eq:spider}
\ena

\noi Another non--Coulomb--like contribution comes from the
two--loop bubble diagram shown in Figure \ref{fig:diagrams}g

\eqa
V^{(g)}(\vR) &=& \frac{g^6}{48} V_{Coul}(\vR ) + V^{(g)}_{nCoul}(\vR )~,
\\
\nonumber \\
V^{(g)}_{nCoul}(\vR ) &=& \frac{g^6}{6N_s^3} \sum_{\vp}
\frac{1 - \cos \vp \vR }{({\vec \cak}^{\, 2})^2 } 
\sum_{i,j=1}^3 \cak_{i} \cak_{j}\cdot 
\Bigl( T_{ij}(\vp )- T_{ij}(\vzero ) \Bigr)~,
\label{eq:big_mac}
\ena

\noi where
\eq
T_{ij}(\vp ) = \frac{1}{(N_4N_s^3)^2} \sum_{q \qr}
D_{ij}(q)  D_{ij}(\qr )  D_{ij}(p-q-\qr ) ~,
\quad
D_{ij}(q) = \frac{ \cak_{i} \cak_{j} +
 \cak_4^2 \cdot \delta_{i j} }{\cak^2}(q) ~.
\en

\noi Therefore, up to order $O(g^6)$ the static charge
potential $V_P(\vR)$ is

\eq
V_{P}(\vR) = g^2_{2-loop} \cdot V_{Coul}(\vR) + g^6 \cdot V_{nCoul}(\vR)~,
\label{eq:VP_pert}
\en
with
\eq
V_{nCoul}(\vR) = V^{(f)}(\vR) + V^{(g)}_{nCoul}(\vR)~,
\label{eq:VnCoul}
\en
and where
\eq
g^2_{2-loop} = g^2 (1+\frac{1}{4}g^2+\frac{11}{96}g^4)~.
\label{eq:g_2l}
\en

\begin{figure}[pt]
\vspace{10.0cm}
\includegraphics{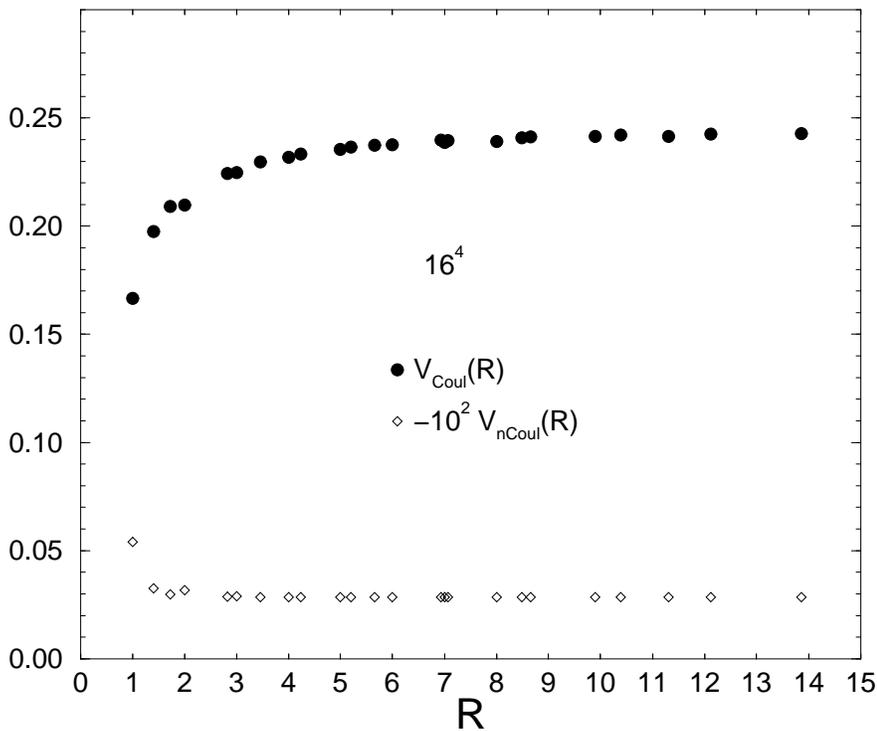}
\caption{Comparison of the non--Coulomb part $V_{nCoul}$,
 eq.~(\protect{\ref{eq:VnCoul}}), of the potential
to the (lattice) Coulomb part $V_{Coul}$,
eq.~(\protect{\ref{v_coul_latt_fin}}), on a $16^4$ lattice.}
\label{fig:nCoul}
\end{figure}

To get an idea of the possible importance of the non--Coulomb term, we
computed it by numerically performing the necessary lattice momentum sums
on several lattices up to size $16^4$.\footnote{To perform the necessary
lattice sums took a couple of weeks CPU time on an IBM RS6000 workstation
for the largest lattice!} As an example we compare
$V_{nCoul}$ with $V_{Coul}$ in Figure \ref{fig:nCoul}. We see that
$V_{nCoul}$ is almost three orders of magnitude smaller than $V_{Coul}$,
a difference that will even be enhanced when including the coupling
constants for weak coupling. Furthermore $V_{nCoul}$ approaches its
asymptotic value much faster than $1/R$. Since the long distance behavior
is governed by the small momentum region, we can estimate it in the
infinite volume limit from power counting in the small momentum region of
the integrals. For the two-loop bubble diagram, eq.~(\ref{eq:big_mac}),
since the zero-momentum part corresponding to $T_{ij}(\vzero)$, which
contributes to $V_{Coul}$, was split off, we expect a $1/R^3$ approach to a
constant. For the four--prong--spider graph, eq.~(\ref{eq:spider}), we
expect, apart from a constant, a $1/R^5$ fall-off at large distance.
Log--log plots of the numerically computed non-Coulomb contributions on
finite lattices confirm these expectations.

From the large distance behavior in the infinite volume limit we can also
estimate the finite size effects. Periodic boundary conditions can be
mimicked by mirror charges, {\it e.g.} for an on-axis distance the infinite
volume $1/R^n$ becomes in a finite volume with box size $N_s$, 
\eq
\frac{1}{R^n} + \frac{1}{(N_s-R)^n} = \frac{1}{R^n} + 
\frac{1}{N_s^n} + n \frac{R}{N_s^{n+1}} + 
\frac{1}{N_s^n} O\left( \frac{r^2}{N_s^2} \right)
\en
Therefore we expect
the leading finite volume corrections to affect the constant at $O(1/N_s)$ 
for $V_{Coul}$ and only at $O(1/N_s^3)$ for $V_{nCoul}$. The finite size
effects in $V_{nCoul}$ appear to be much smaller than those in $V_{Coul}$.
Our numerical results are in agreement with this expectation.

\section{The potential and the renormalized coupling from Wilson loops}

The potential can also be obtained from Wilson loops: $V = V_W(T,\vR )$.
It is defined as follows

\eq
V_{W}(T;{\vec R}) = \ln \frac{W(T,{\vec R})}{W(T+1,{\vec R})}~,
\en

\noi where $~W(T,\vR)~$ is the Wilson loop with `time'
extension $T$ and space extension $\vR$.

The Wilson loop can be 
on--axis or off-axis, and the space--like parts of the loop
can include the contribution of many different contours.
In our calculations we have chosen planar loops and
two types of non--planar contours: the `plane-diagonal' contour
with space--like part in the plane $(x_2,x_3)$ 
at fixed $x_1$ (say, $x_1 = 0$) connecting
points $x_2 = x_3 = 0$ and $x_2 = x_3 = R_0$, and
the `space-diagonal' contour in the $3d$ space $(x_1,x_2,x_3)$ 
connecting points $x_1 = x_2 = x_3 = 0$ and 
$x_1 = x_2 = x_3 = R_0$. For the off-axis loops we average over all
paths that follow the straight line between the endpoints as closely
as possible. For example, for the loop through the origin and
$x_2 = x_3 = R_0$ the paths considered go through all points with
$x_2 = x_3$ in between. We average over all combinations of first taking
a step in the 2-direction followed by a step in the 3-direction and
vice versa.

At lowest order in perturbation theory the potential $V_W(T;\vR )$ becomes

\eq
V_W(T;\vR) = w(T,\vR) - w(T+1,\vR)~,
           \label{lowest_order_a}
\en

\noi with $w(T,\vR)$ the $O(g^2)$ part of the Wilson loop, which in turn, in
Feynman gauge, can be split as
\eq
w(T,\vR)  = w_s(T,\vR) + w_t(T,\vR)~,
\en

\noi where $~w_t~$ and $~w_s~$ are contribution of the time--time
(`electric') and space--space (`magnetic') parts, respectively.
The time--time contribution is

\eq
w_t(T,\vR) = -\frac{g^2}{N_4N_s^3 } \sum_{p}
( 1 - \cos \vp \vR ) \cdot
\frac{1 - \cos p_4T}{1 - \cos p_4} \cdot G_0(p)~,
\en

\noi where $G_0(p)=1/\cak^2$.
It is easy to show that in the limit
$~N_4 \ra \infty~,~~T \ra \infty~,~~T \ll N_4~$
the 'time'--like ('electric')
part of the potential gives the lattice Coulomb potential as obtained from
Polyakov loop correlations and given in eq.~(\ref{v_coul_latt_fin}).

For the planar Wilson loops in the plane $~(x_3,x_4)~$
the space--space contribution $~w_s~$ is

\eq
w_s = -\frac{g^2}{N_4N_s^3 } \sum_{p\ne 0}
(1 - \cos p_4T ) \cdot \frac{1 - \cos p_3R }{1 - \cos p_3 }
\cdot G_0(p)~.
\en

\noi For the `plane-diagonal' contour, in the plane $~(x_2,x_3)~$ 
at fixed $~x_1~$ connecting points $~x_2 = x_3 = 0~$ with $~x_2 = x_3 = R_0~$
the space--space contribution is

\eq
w_s = -\frac{g^2}{2N_4N_s^3} \sum_{p\ne 0}
(2 + \cos p_2 +\cos p_3) \cdot 
\frac{1 - \cos R_0(p_2+p_3) }{1 - \cos (p_2+p_3) } \cdot 
( 1 - \cos p_4T ) \cdot G_0(p )  ~,
\en

\noi and finally for the `space-diagonal' contour
connecting points $~x_1 = x_2 = x_3 = 0~$ with 
$~x_1 = x_2 = x_3 = R_0~$ the space--space contribution is

\eqa
w_s &=& \frac{2g^2}{N_4N_s^3} \sum_{p\ne 0} 
{\cal A}(\vp ) 
\cdot \frac{1 - \cos R_0(p_1+p_2+p_3) }{1 - \cos (p_1+p_2+p_3) } 
\cdot ( 1 - \cos p_4T ) \cdot G_0(p) ~,
\nonumber \\
\nonumber \\
{\cal A} &=& -\frac{1}{18}
\left[~2 \cos \frac{p_2+p_3}{2} + \cos \frac{p_2-p_3}{2} ~\right]^2
+ \Bigl\{ \mbox{permutations of} ~p_1,p_2,p_3 \Bigr\}~.
           \label{lowest_order_b}
\ena

The contribution from $w_s$ to $V_W(T;\vR)$ vanishes in the limit
$T \to \infty$ and in this limit the potentials extracted from Polyakov
line correlations and from Wilson loops agree. At finite $T$, however,
$w_s$ gives a contribution $\approx c R/T^2$ to $V_W(T;\vR)$. At a
finite $T$ the potential $V_W(T;\vR)$ hence appears to have a
small confining contribution. $V_W(T;\vR)$ therefore needs to be either
carefully extrapolated to infinite $T$, or the finite $T$ effect has to
be taken into account in the analysis, as we shall do in this paper.

Just as was the case for the potential obtained from Polyakov loop
correlations, at one--loop level the only effect is a
renormalization of the coupling $g^2 \to g^2 (1 + \frac{1}{4}g^2)$.
At two loops in the perturbative expansion we only considered, for
simplicity, planar Wilson loops. We find the same structure as for
the potential from Polyakov loops, eq.s~(\ref{eq:VP_pert}) and
(\ref{eq:VnCoul}). In particular, the two--loop renormalization of the
coupling is identical, as given in eq.~(\ref{eq:g_2l}).

Our Monte Carlo data were produced on a $~16^3\times 32~$ lattice for
$11$ values of $~\beta~$ shown in Table ~\ref{tab:g_ren}.  We measured
planar and nonplanar Wilson loops $~W(T,\vR )~$ with $~T\le T_{max}=15~$
and $~R\le R_{max}=8\sqrt{3}$. To decrease the statistical noise we have used
the Parisi--Petronzio--Rapuano trick \cite{ppr} for all time--like links.
Because of this we do not have a valid measurement of the potential at
distance $R=1$, and this distance is therefore excluded from all further
considerations.

\begin{figure}
\vspace{10.0cm}
\includegraphics{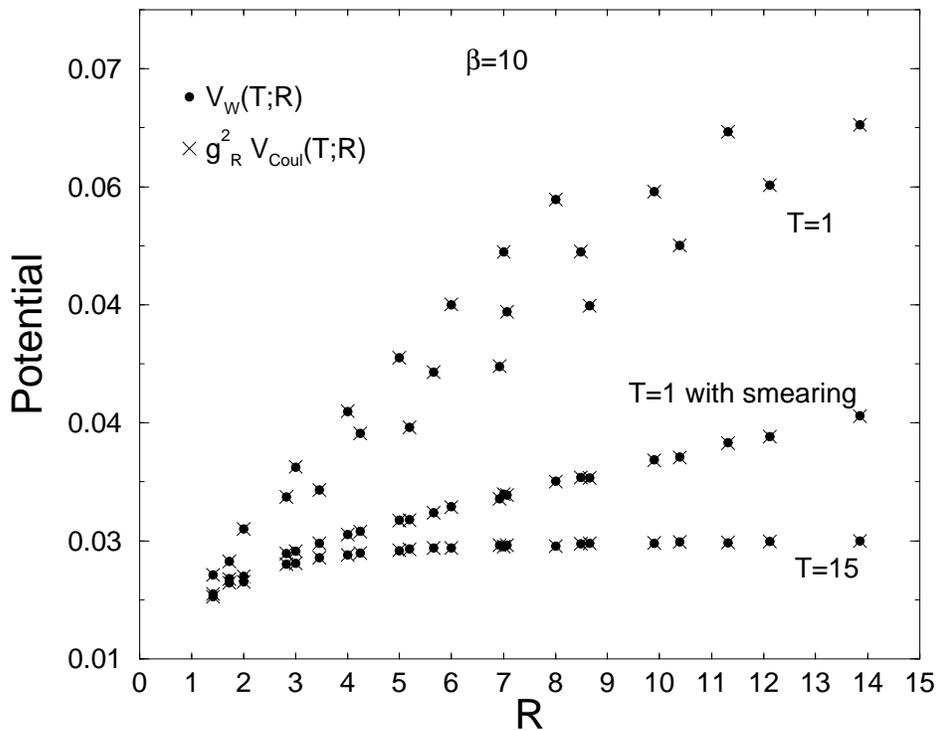}
\caption{The Wilson loop potential $V_W(T;\vR )$ (circles) 
and the corresponding perturbative potential $g^2_RV_{Coul}(T;\vR )$ (crosses)
at $\beta =10$ on a $16^3 \times 32$ lattice with and without smearing. 
For $T=15$ the potential from smeared and normal Wilson loops would be
indistinguishable in the figure.
}
\label{fig:potential}
\end{figure}

In the Coulomb phase the potential obtained from Monte Carlo simulations is
expected to be very similar to the perturbative lattice Coulomb potential.
Since the latter can have sizable finite size and finite $T$ effects, as
discussed earlier, we decided to use as a fit formula for the numerical
potential $V_W(T;\vR )$ the lowest order perturbative expression defined in
eq.'s~(\ref{lowest_order_a}) to (\ref{lowest_order_b}), with the only fit
parameter being the renormalized coupling constant $g^2_R$. The success of
such fits will indicate that the non-Coulomb contributions are small, like
in perturbation theory, and that non-perturbative effects just go into the
renormalization of the coupling. The number of degrees of freedom,
$N_{d.o.f.}$ in our fits varied between $N^{max}_{d.o.f.}=344$ and
$N^{min}_{d.o.f.}=31$ with decreasing $\beta$, from $\beta=10$ to
$\beta=1.015$. In all the cases we obtained $\chi^2/N_{d.o.f.}  \aleq 1.0$
(with the only exception being for $\beta=1.015$ where $\chi^2/N_{d.o.f.}
\simeq 2.1$). As an example we show in Figure \ref{fig:potential} the
$R$--dependence of the potential $V_W(T;\vR )$ at $T=1$ and $T=15$
(circles).   Crosses show the corresponding values of $g^2_RV_{Coul}(T;\vR
)$ with the fitted $g^2_R$. We would like to emphasize that, even at
$T=15$, the potential from nonplanar Wilson loops cannot be fitted with a
continuum--like $1/R$ ansatz, which can be attempted when only planar Wilson
loops are available (see {\it e.g.} \cite{jnz,sw1}).

At large values of $\beta$ ($\beta \ge 5$) our one-parameter fit works well
for the whole set of data points, {\it i.e.} all points with $1\le T\le 15$
and $1<R\le 8\sqrt{3}$ ($N_{d.o.f.}=344$). The non--Coulomb--like
corrections are therefore negligible despite our rather small statistical
errors. With decreasing $\beta$ these corrections become noticeable at
small values of $R$ (recall the fast fall-off of $V_{nCoul}$ like $1/R^3$)
and small values of $T$. After excluding the data points corresponding to
these small $R$ and $T$ values from the fit, we obtained fits with a high
confidence level, but fewer degrees of freedom, as described above.
Remarkably, the lattice Coulomb potential works, at least at large
distance, really well even at $\beta$--values very close to the phase
transition point, $\beta^\ast$, to the confined phase. This observation in
fact constitutes one of the main results of this paper.

\begin{figure}
\vspace{10.0cm}
\includegraphics{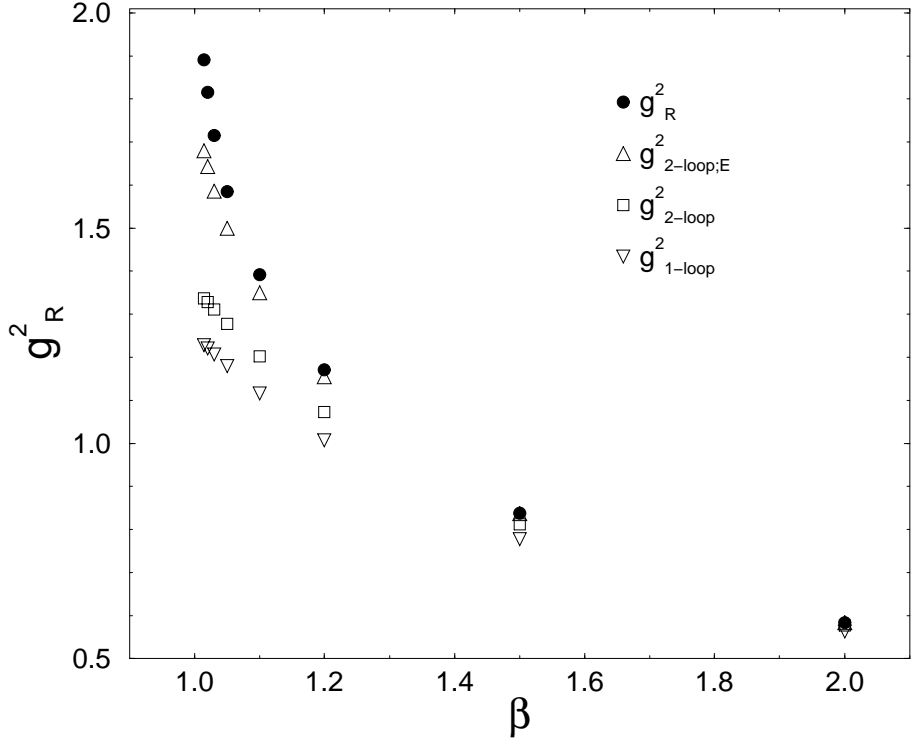}
\caption{The renormalized coupling $g^2_R$ extracted from the fit on a 
$16^3 \times 32$ lattice as compared with $1$--loop and $2$--loop perturbative
values.}
\label{fig:g_ren}
\end{figure}

The extracted values of the renormalized coupling $g^2_R$ as well as
the perturbative values $g^2_{1-loop}$ and $g^2_{2-loop}$ are
listed in Table \ref{tab:g_ren} and shown in Figure \ref{fig:g_ren}.
The difference between $g^2_R$ and the perturbative value
$g^2_{2-loop}$ becomes noticeable at $\beta \aleq 2.0$.
This is the region where one expects the perturbative three--loop
contributions to become important. In addition nonperturbative effects may
start to significantly contribute.

\begin{table}[htb]
\begin{center}
\begin{tabular}{|c|c|c|l|} \hline
\multicolumn{1}{|c|}{$\beta$} & \multicolumn{1}{|c|}{$g^2_{1-loop}$}
&\multicolumn{1}{|c|}{$g^2_{2-loop}$} &\multicolumn{1}{|c|}{$g^2_R$} \\ \hline\hline
  1.015 &  1.22789  & 1.33747  & 1.892(5)        \\ \hline
  1.020 &  1.22068  & 1.32866  & 1.815(5)        \\ \hline
  1.030 &  1.20652  & 1.31138  & 1.716(4)        \\ \hline
  1.050 &  1.17914  & 1.27812  & 1.585(2)        \\ \hline
  1.100 &  1.11570  & 1.20179  & 1.3913(10)      \\ \hline
  1.200 &  1.00694  & 1.07325  & 1.1707(7)       \\ \hline
  1.500 &  0.77778  & 0.81173  & 0.8379(5)       \\ \hline
  2.000 &  0.56250  & 0.57682  & 0.5833(3)       \\ \hline
  5.000 &  0.21000  & 0.21092  & 0.21104(4)      \\ \hline
  7.000 &  0.14796  & 0.14829  & 0.14834(3)      \\ \hline
  10.00 &  0.10250  & 0.10262  & 0.10263(3)      \\ \hline
\end{tabular}
\end{center}
\caption{The renormalized coupling $g^2_R$ extracted from the fit on a 
$32\times 16^3$ lattice as compared with $1$--loop and $2$--loop perturbative
values.
}
\label{tab:g_ren}
\medskip \noindent
\end{table}

In numerical simulations of QCD it has become standard to use a
smearing procedure to improve the signal--to--noise ratio. In the
computation of Wilson loops the space--like link elements $U_k$
are replaced be recursively computed ``smeared links'' $U^{(n)}_k$

\eq
U^{(n+1)}_{xk} = P \Bigl[ \alpha U^{(n)}_{xk} + \sum_{j \ne k}
 \bigl( U^{(n)}_{xj} U^{(n)}_{x+j,k} U^{(n)\dagger}_{x+k,j}
+ U^{(n)\dagger}_{x-j,j} U^{(n)}_{x-j,k} U^{(n)}_{x-j+k,j}
\bigr) \Bigr], \quad j,k=1,2,3~,
\label{eq:APE}
\en

\noi where $P$ indicates a projection onto the gauge group. The choice
of $\alpha$, the weight of the direct link relative to the ``staples'',
and the number of smearing iterations $n_{sme}$ are parameters that can
be optimized.

At lowest perturbative order the recursive
relation eq.~(\ref{eq:APE}) becomes for Fourier transformed fields

\eq
A^{(n+1)}_k(p) = \frac{1}{\alpha+4} \Bigl[ A^{(n)}_k(p) \bigl( \alpha 
+ \sum_{j \ne k} 2\cos p_j \bigr) 
+ 4 \sin\frac{p_k}{2} \sum_{j \ne k} A^{(n)}_j(p) \sin\frac{p_j}{2} \Bigr].
\en

\noi This relation can easily be iterated and inserted in the numerical
computation of the momentum sums to obtain the lowest order perturbative
contribution to the potential from the space--space parts of the
Wilson loops.

In Figure \ref{fig:potential} we show the effect of smearing on the
potential extracted at $T=1$ (circles) for $\beta=10$. The parameters of
smearing were chosen as $\alpha=4$ and $n_{sme}=6$. The crosses on the top
of these circles indicate the values of the potential obtained from the
one-parameter fit to the perturbative form, as described above. 
In this case the $\chi^2$--value is somewhat larger 
($\chi^2/N_{d.o.f.} \sim 2.9$). However the extracted value of the 
renormalized coupling $g^2_R$ appears to be very stable with respect
to the smearing procedure, and $g^2_R$ from the smeared loops 
coincides (within errorbars) with the coupling found without smearing.

For non--abelian lattice gauge theory it has proved useful, in perturbative
computations, to use an the `improved' coupling, such as $g^2_E$ 
\cite{par1,par2}

\eq
g^2_E \equiv \frac{1}{c_1} \Bigl(1 - \langle U_p \rangle \Bigr)~,
               \label{g_E}
\en

\noi where $U_p$ denotes the plaquette and $c_1=1/4$ is the first
coefficient in its perturbative expansion. For pure gauge $U(1)$ the
perturbative expansion of $\langle U_p \rangle$ is known to three
loops\cite{H_W}\footnote{We have checked the two--loop coefficient
({\it i.e.} $\sim g^6$) which we use below}

\eq
\langle U_p \rangle = 1 - \frac{1}{4} g^2 - \frac{1}{32} g^4 -
 0.0131185 g^6 - 0.00752 g^8 +\ldots~.
\en

\noi From this we get a perturbative expansion of $g^2_E$ in terms of the
bare $g^2$. We then can express $g^2$ in terms of $g^2_E$ and substitute
in the perturbative formula for $g^2_{2-loop}$, eq.~(\ref{eq:g_2l}),

\eq
g^2_{2-loop,E} = g^2_E (1 + \frac{1}{8} g^2_E + 0.0308594 g^4_E +\ldots ).
\label{eq:gE_2l}
\en

\noi The plaquette averages, $g^2_E$ and the corresponding prediction for
$g^2_{2-loop,E}$ from eq.~(\ref{eq:gE_2l}) is listed in Table
\ref{tab:g_E}. Comparing with Table \ref{tab:g_ren} we see, that
$g^2_{2-loop,E}$ based on the `improved' coupling $g^2_E$ is indeed a
somewhat better prediction for $g^2_R$ than the purely perturbative
$g^2_{2-loop}$. This fact can also be seen in Figure \ref{fig:g_ren}.

\begin{table}[htb]
\begin{center}
\begin{tabular}{|c|c|c|c|} \hline
  $\beta$      &    $\langle U_p \rangle$ & $g^2_E$ & $g^2_{2-loop,E}$
                                                         \\ \hline \hline
  1.015 &   0.66147(2)    &  1.3541  &  1.6800   \\ \hline
  1.02  &   0.66749(3)    &  1.3301  &  1.6429   \\ \hline
  1.03  &   0.67680(2)    &  1.2928  &  1.5859   \\ \hline
  1.05  &   0.69104(2)    &  1.2358  &  1.5002   \\ \hline
  1.1   &   0.71674(2)    &  1.1330  &  1.3501   \\ \hline
  1.2   &   0.751625(10)  &  0.9935  &  1.1551   \\ \hline
  1.5   &   0.812599(6)   &  0.7496  &  0.8363   \\ \hline
  2     &   0.864810(8)   &  0.5408  &  0.5835   \\ \hline
  5     &   0.9486330(24) &  0.20547 &  0.21109  \\ \hline
  7     &   0.9636091(13) &  0.14556 &  0.14833  \\ \hline
 10     &   0.9746754(8)  &  0.10130 &  0.10262  \\ \hline
\end{tabular}
\end{center}
\caption{The average plaquette, $\langle U_p \rangle$, the extracted
$g^2_E$ and the corresponding prediction for $g^2_{2-loop,E}$ from
eq.~(\protect\ref{eq:gE_2l}) 
}
\label{tab:g_E}
\medskip \noindent
\end{table}

\section{Universal maximal coupling?}
\label{max_coup}

In paper \cite{car} the existence of a universal finite value $4\pi
\alpha_c \equiv g^2_c$ was predicted at the deconfinement point,
$\beta^\ast$, based on an analogy with the $2d$ Kosterlitz--Thouless
transition. The mechanism of the transition in the pure gauge $U(1)$
theory and in the $2d$ $XY$ model was shown to be the termination of
the massless phase by a topological disordering. The approach to this
maximum finite coupling was conjectured as

\eq
 g^2_R = g^2_c - c \left( \beta - \beta^\ast \right)^\lambda
\label{eq:g2c_form}
\en

\noi with the estimates $g^2_c =4\pi \alpha_c = 1.90\pm 0.10$ and
$\lambda = 0.5(1)$ \cite{luck2}.

We have attempted fits of the form eq.~(\ref{eq:g2c_form}) to our data.
A four--parameter fit to the 5 data points with the smallest $\beta$
gives $\beta^{\ast}=1.007(3)$, $g^2_c=2.4(3)$ and $\lambda=0.27(8)$ with a
$\chi^2 = 0.15$ for 1 degree of freedom. Including more data points
decreases the estimate of $\beta^\ast$ and increases the estimate of
$g^2_c$. The above estimate for $\beta^\ast$ is somewhat smaller then
the value $\beta^\ast = 1.011(2)$ found in \cite{jnz}. Constraining
$\beta^\ast$ to 1.011 a three--parameter fit gives now $g^2_c=2.12(2)$ and
$\lambda=0.38(2)$ with a $\chi^2=1.20$ for 2 degrees of freedom, or,
using only 4 data points, $g^2_c=2.08(4)$ and $\lambda=0.43(5)$ with a
$\chi^2=0.01$ for 1 degree of freedom. Varying $\beta^\ast$ from 1.009
to 1.013 gives variations somewhat larger than the statistical errors.
Our final estimates are then

\eq
 g^2_c = 2.08 \pm 0.14 ~~~~~~~ {\rm and} ~~~~~~~ \lambda = 0.43 \pm 0.10
\label{eq:g2c}
\en

\noi with the errors dominated by systematic uncertainties.
Obviously, more data points in the ``critical'' region would be desirable
to make the systematic errors smaller.

The existence of $g^2_c$ as maximal coupling depends on the presence of
the monopole induced phase transition to a confining phase. When the
monopoles are suppressed, this phase transition disappears. This leads to
the question of whether in that case larger renormalized couplings can be
achieved than the above $g^2_c$. This seems indeed the case as found in
\cite{KHMM} in the context of the U(1) Higgs model with suppressed
monopoles. We have repeated the measurements of the renormalized coupling
for pure U(1) with completely suppressed monopoles on an $8^3 \times 16$
lattice. The results are listed in Table \ref{tab:g_R_NM}. $g^2_R$ was
extracted exactly as for the model with standard Wilson action described in
the previous section. Since our analysis carefully includes finite size
(and finite $T$) effects, a lattice of size $8^3 \times 16$ is sufficient
for our purposes here. In Table \ref{tab:g_R_NM} we list, for comparison,
also renormalized couplings for the standard model obtained from an
$8^3 \times 16$ lattice. Comparison with Table \ref{tab:g_ren} shows that
the residual finite size effects are indeed very small, even close to the
phase transition.

\begin{table}[htb]
\begin{center}
\begin{tabular}{|c|c|c|}
\hline
 $\beta$ & $g^2_R$(NoMo) & $g^2_R$(stand) \\ \hline
 0.0001  & 5.19(4)   &            \\ \hline
 0.25    & 3.97(2)   &            \\ \hline
 0.5     & 2.911(11) &            \\ \hline
 1.0     & 1.445(5)  &            \\ \hline
 1.015   &           & 1.86(3)    \\ \hline
 1.02    & 1.406(3)  & 1.79(3)    \\ \hline
 1.05    & 1.349(3)  & 1.579(8)   \\ \hline
 1.2     & 1.125(2)  & 1.172(2)   \\ \hline
 1.5     & 0.8315(8) & 0.8362(13) \\ \hline
 2.0     & 0.5824(5) & 0.5831(2)  \\ \hline
\end{tabular}
\end{center}
\caption{The coupling $g^2_R$ from an $8^3\times16$ lattice for the
standard U(1) model, and for the model with complete monopole suppression.}
\label{tab:g_R_NM}
\medskip \noindent
\end{table}

From Table \ref{tab:g_R_NM} we see that at larger $\beta$ ($\beta \geq 2.0$)
monopole suppression has no effect, as expected, since monopoles are very
much suppressed in the standard U(1) model there. For smaller $\beta$, due
to the effect from monopoles, $g^2_R$ raises faster for the standard U(1)
model until reaching its maximal value $g^2_c$ at the phase transition
point. With monopoles completely suppressed, $g^2_R$ raises more slowly.
However, the absence of the confinement phase transition, allows $g^2_R$ to
keep growing and to become larger than $g^2_c$. Although the model with
complete suppression of monopole has no phase transition even at $\beta=0$,
we stopped our measurements at a small positive value to ensure that the
bare coupling remains real.

The maximal coupling $g^2_c$, however, should be universal
for U(1) actions with a monopole
induced phase transition, for example the Villain model. The renormalized
coupling was measured for the Villain model in \cite{dgt1}, albeit only on
a $6^3 \times 4$ lattice. We simulated an $8^3 \times 16$ lattice and
extracted $g^2_R$ the same way as for the Wilson action. Our results are
listed in Table \ref{tab:g_R_Vil}.

\begin{table}[htb]
\begin{center}
\begin{tabular}{|c|c|}
\hline
 $\beta$ & $g^2_R$   \\ \hline
 0.645   & 2.027(30) \\ \hline
 0.65    & 1.915(24) \\ \hline
 0.655   & 1.868(12) \\ \hline
 0.66    & 1.816(9)  \\ \hline
 0.67    & 1.732(6)  \\ \hline
 0.68    & 1.672(5)  \\ \hline
 0.7     & 1.575(6)  \\ \hline
 0.75    & 1.402(2)  \\ \hline
 0.8     & 1.285(2)  \\ \hline
 1.0     & 1.003(2)  \\ \hline
\end{tabular}
\end{center}
\caption{The coupling $g^2_R$ from an $8^3\times16$ lattice for the
U(1) model with Villain action.}
\label{tab:g_R_Vil}
\medskip \noindent
\end{table}

For the Villain action, there is no perturbative renormalization of the
bare coupling. All the renormalization comes from the monopoles. Indeed,
up to $\beta=1.0$ the renormalized and bare couplings are almost identical.
Only then do the monopoles start to have an appreciable influence and
$g^2_R$ becomes larger than the bare coupling and continues to raise until
the phase transition to the confined phase is reached. We made a fit of our
data to the form eq.~(\ref{eq:g2c_form}). The fit works very nicely for
$\beta \leq 0.75$. Including all 8 data points gives a $\chi^2 = 1.23$ for
4 degrees of freedom with

\eq
 g^2_c = 2.12(4) ~,~~~ \beta^\ast = 0.6457(8) ~~~{\rm and} ~~~
 \lambda = 0.43(4)
\label{eq:g2c_Vil}
\en

\noi in nice agreement with the results for the Wilson action,
eq.~(\ref{eq:g2c}) and confirming the universality of $g^2_c$. The result
eq.~(\ref{eq:g2c_Vil}) is stable when omitting data points from the fits.

\section{Conclusions}

We have made an analytical and numerical study of compact lattice pure
gauge $~U(1)~$ theory in the Coulomb phase. The main point of interest
was the study of the {\it non}--Coulomb contributions to the heavy charge
potential. For this purpose we calculated perturbatively the heavy charge
potential $~V_P({\vec R})~$ defined from the Polyakov loop correlations in
the 2--loop (i.e., $\sim g^6$) approximation. This calculation shows that
the non--Coulomb contribution $~V_P^{nCoul}(\vR)~$ is much smaller than the
Coulomb contribution $~V_P^{Coul}(\vR)~$, at least at large distances.

The conclusions obtained within perturbation theory were confirmed by
numerical calculations  of the potential $~V_W(T;\vR)~$ defined from
planar and nonplanar time--like Wilson loops.  We used as a fit formula for
$~V_W(T;\vR)~$ the lowest order perturbative expression with the only fit
parameter being the renormalized coupling constant $~g^2_R$. These fits
worked well for all distances at large $\beta$, and for sufficiently big
$~T~$ and $~R$ even down to very close to the phase transition to the
strong coupling confined phase.
It is worth noting that it is impossible to obtain a good $~\chi^2~$
using for the fit a continuum potential $~\sim 1/R$. The values of the
renormalized coupling $~g^2_R~$ are stable with respect to a smearing
procedure.

We confirmed, with better accuracy, the conjectured existence of a
universal maximal coupling for $U(1)$ models with a monopole induced
confining phase transition. In the model with complete suppression of
monopoles, however, even larger renormalized couplings can be reached.

Our main conclusion is that {\it compact} pure gauge $U(1)$
theory can serve equally well as a non--compact version to describe
the physics of free photons in the weak coupling region.
The only difference is a finite, $R$--independent
renormalization of the coupling constant  $~g^2\to g^2_{R}$ in the compact
theory.

\end{document}